\documentclass[a4paper,12pt]{article}
\usepackage{bm, color}
\usepackage{amsmath,amsfonts,amssymb}
\usepackage{graphicx}
\usepackage{amsmath,amsfonts,amssymb,bm}
\usepackage{color}
\usepackage{subfigure}
\usepackage{placeins}
\begin{document}
\title{An efficient GFET structure}
\date{}
\author{Giovanni Nastasi\thanks{Department of Mathematics and Computer Science, Universit\`{a} degli Studi di Catania, viale Andrea Doria 6, 
95125 Catania, Italy  ({\tt g.nastasi@unict.it}).} 
\and Vittorio Romano\thanks{Department of Mathematics and Computer Science, Universit\`{a} degli Studi di Catania, viale Andrea Doria 6, 
95125 Catania, Italy  ({\tt romano@dmi.unict.it}).} }
\maketitle

\begin{abstract}
A graphene field effect transistor, where the active area is made of monolayer large-area graphene, is simulated including a full 2D Poisson equation and a drift-diffusion model with mobilities deduced by a direct numerical solution of the semiclassical Boltzmann equations for charge transport by a suitable discontinuous Galerkin approach. 

The critical  issue in a graphene field effect transistor is the difficulty of fixing the off state which requires an accurate calibration of the gate voltages. In the present paper we propose and simulate a graphene field effect transistor structure which has well-behaved characteristic curves similar to those of conventional (with gap) semiconductor materials. The introduced device  has a clear off region and can be the prototype of  devices  suited for post-silicon nanoscale electron technology.
The specific geometry overcomes the problems of triggering the minority charge current and gives a viable way for the design of electron devices based on large area monolayer graphene as substitute of standard semiconductors in the active area. The good field effect transistor  behavior of the current versus the gate voltage makes the simulated device very promising and a challenging case for experimentalists.  
\end{abstract}


\section{Introduction}
The Metal Oxide Semiconductor Field Effect Transistor (MOSFET) is the backbone of the modern integrated circuits. In the case  the active area is made of traditional materials like, for example, silicon or gallium arsenide, a lot of analysis and simulations have been performed in order to optimize the design. Lately a great attention has been devoted to graphene  on account of its peculiar features, and in particular, from the point of view of nano-electronics, for the high 
electrical conductivity. It is highly tempting to try to replace the traditional semiconductors with graphene in the active area of electron devices like the MOSFETs \cite{Sch}   even if many aspect  about the actual performance  in real applications
remain unclear.

Scaling
theory predicts that a FET with a thin barrier and a thin gate-controlled
region will be robust
against short-channel effects down to very short gate lengths. The possibility of having
channels that are just one atomic layer thick is perhaps the most
attractive feature of graphene for its use in transistors.
Main drawbacks of a large-area single monolayer graphene are the  zero gap and, for graphene on substrate, the degradation of the mobility. Therefore accurate simulation are warranted for the set up of a viable
 graphene field effect transistor.
 
The standard mathematical model is  given by the drift-diffusion-Poisson system.
Usually the GFETs are investigated by adopting reduced one dimensional models of the Poisson equation with some averaging procedure \cite{Jimenez,Upadhyay,Meric}. Here a full two-dimensional simulation is presented.  

A crucial point is the determination of the mobilities entering the drift-diffusion equations. A rather popular model is that proposed in \cite{Dor}. Here a different approach is adopted. Thanks to the discontinuous Galerkin (DG) scheme developed in \cite{RoMajCo,CoMajRo,ART:CoMajNaRo_AAPP,ART:MajNaRo_CICP}, we have performed, for graphene on a substrate, an extensive numerical simulation based on the semiclassical Boltzmann equations, including electron-phonon and electron-impuritiy scatterings along with the scattering with remote phonons of the substrate. Both  intra and inter-band scatterings  have been taken into account. For other simulation approaches the interested reader is referred to \cite{LMS,MuWa}. Quantum effects have been also introduced in \cite{Bar,Morandi11,LuRo1,MaRo2,MaRo_entropy}. An alternative approach could be resorting to hydrodynamical models 
(see \cite{CaRo,LuRo2}).

From the numerical solutions of the semiclassical Boltzmann equation a model for the mobility functions has been deduced, similarly to  what already done in \cite{MajMaRo} and in \cite{ART:NaRo_CAIM} in the case of suspended monolayer graphene. 

In the present paper we propose a slight different geometry of GFET which leads to a clear and sizable off region, avoiding the problem of triggering the minority charge carriers and producing characteristic curves which seem very promising for the use of large-area graphene  in electron devices.

The plan of the paper is as follows. In Sec. 2 the structure of the proposed device is presented along with the  mathematical model. In Sec. 3 the mobilities are sketched  and in Sec. 4  the numerical simulations of the proposed GFET are shown. 

\section{Device structure and mathematical model}
\begin{figure}[!t]
\centerline{\includegraphics[width=0.8\columnwidth]{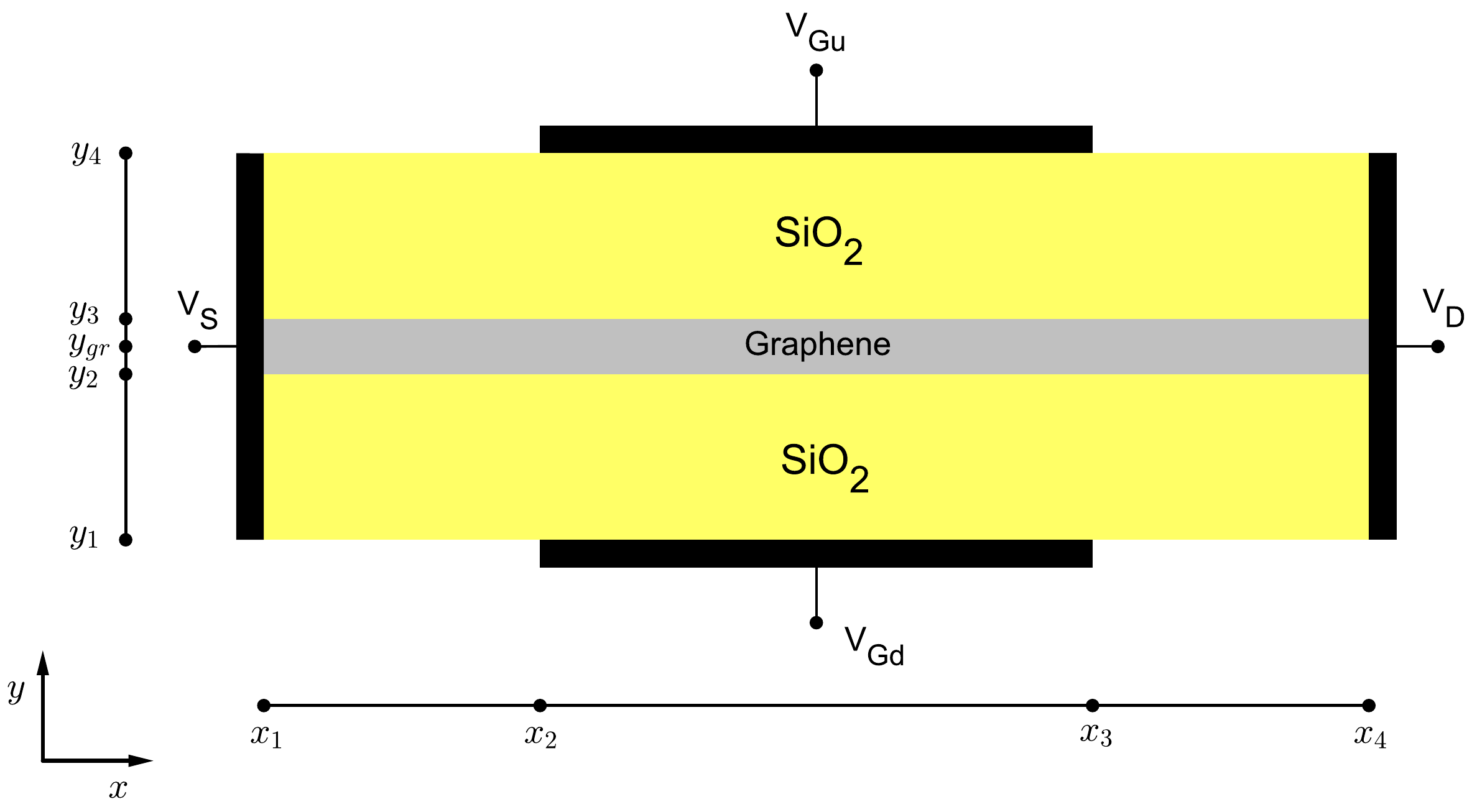}}
\caption{Schematic representation of the GFET investigated in the paper.}\label{FIG:GMOSFET}
\end{figure}

We propose the device with the geometry  depicted in Fig.~\ref{FIG:GMOSFET}. The active zone is made of a single layer of graphene which is between two strips of insulator, both of  them being  SiO$_2$. The source and drain contacts are directly attached to the graphene. The two gate contacts (up and down) are attached to the oxide.  The choice of the type of oxide is not crucial at this stage. We have considered SiO$_2$ as an example. Instead it is crucial to put the source and drain contacts along all the lateral edges for a better control of the electrostatic potential as will be clear from the simulations. In the direction orthogonal to the section the device is considered as infinitely long. 

We solve a 2D Poisson equation for the electrostatic potential, assuming that the charge is concentrated on the volume occupied by the atoms composing the graphene layer. Accordingly, the surface charge density of graphene is supposed to be spanned on its thickness which experimental measurements refer between 0.4 nm and 1.7 nm \cite{Shearer}. In order to simulate the current flowing in the channel we adopt the 1D bipolar drift-diffusion model, coupled to the Poisson equation for the electrostatic potential in the whole section. A special attention is required by the initial carrier density profiles that have to be determined compatibly with the electric potential, leading to a nonlinear Poisson equation as  discussed in the following.
The bipolar drift-diffusion model is adopted in $[x_1,x_4]\times\lbrace y_{gr}\rbrace$ and it reads
\begin{eqnarray*}
\frac{\partial n}{\partial t} - \frac{\partial}{\partial x}\left( \mu_n U_T \frac{\partial n}{\partial x}- n\mu_n\frac{\partial \phi}{\partial x} \right)=0,\\
\frac{\partial p}{\partial t} + \frac{\partial}{\partial x}\left( -\mu_p U_T\frac{\partial p}{\partial x}- p\mu_p\frac{\partial \phi}{\partial x} \right)=0,
\end{eqnarray*}
where $n(t,x)$, $p(t,x)$ are the electron and hole densities in  graphene respectively, at time $t$ and position $x$, $U_T=k_B T/e$ is the thermal voltage, being $e$ the positive elementary charge, $k_B$ is the Boltzmann constant, $T$ is the lattice temperature (kept constant). The functions $\mu_n$ and $\mu_p$ are the mobilities for electrons and holes respectively and $\phi(x,y)$ is the electric potential, here evaluated on $y=y_{gr}$, $y_{gr}$ being  the average $y$-coordinate of the graphene sheet (see Fig. \ref{FIG:GMOSFET}). The generation and recombination terms are set equal to zero \cite{ART:NaRo_CAIM}. Indeed, this relation is strictly valid at steady state but here will be assumed during the transient as well. We expect that the stationary solutions is not affected by such an approximation.  A typical behavior of the total generation and recombination term versus time, obtained with the DG method \cite{RoMajCo,CoMajRo,ART:CoMajNaRo_AAPP,ART:MajNaRo_CICP}, can be found in \cite{NaRo_DGFET}.

The system is solved in the interval $[x_1,x_4]$ augmented with Dirichlet boundary conditions for electron and hole densities as
will be explained below. 

 The electric potential solves the 2D Poisson equation 
\begin{equation}\label{EQ:Poiss}
\nabla\cdot(\epsilon\nabla\phi)=h(x,y),
\end{equation}
where
$h(x,y) = e(n(x)-p(x))/t_{gr}$ if $\, (x, y) \in [x_1, x_4] \times [y_2, y_3] $ and 
$h(x,y) = 0$ otherwise,
and $\epsilon$ is given by
$\epsilon_{gr} $  if $y \in [y_2, y_3]$ and
$\epsilon_{ox}$ otherwise.
Here $\epsilon_{gr} = 3.3 \, \epsilon_0$ and $\epsilon_{ox} 3.6 \, \epsilon_0$ are the dielectric constants of the graphene and oxide (SiO$_2$) respectively,  $\epsilon_0$ being the dielectric constant in the vacuum;
 $t_{gr}$ is the width of the graphene layer 
which is assumed to be 1 nm (values between 0.4 and 1.7 nm are reported in \cite{Shearer}).  The charge in the graphene layer is considered as distributed in the volume enclosed by the parallelepiped of base the area of the graphene  and height $t_{gr}$. Recall that $n$ and $p$ are areal densities. Dirichlet conditions are imposed on the  gate contacts and homogeneous Neumann conditions on the external oxide edges. 
A major issue is to model the source and drain regions where metal and graphene touch. We assume that source and drain are thermal bath reservoirs of  charges which obey a Fermi-Dirac distribution.  The injection of charges is determined by a work function $W_F$.  Indeed it depends on the specific material the contacts are made of. 
We set $W_F = 0.25 V$ which is appropriate for Cu within the experimentally reported range of 0.20 eV  \cite{FraVe} and 0.30 eV  \cite{WaNi}.

As summary the following boundary conditions for the electric potential  are imposed
\begin{eqnarray*}
\begin{array}{ll}
\phi = V_S = W_F & \quad \mbox{at} \quad  x= x_1, \, y \in [y_1, y_4]\\
\phi = V_D = W_F  + V_b & \quad \mbox{at} \quad  x= x_4, \, y \in [y_1, y_4]\\
\phi = V_{Gu} = W_F + V_{G} &\quad \mbox{at} \quad y=y_4, \, x \in [x_2, x_3]\\
\phi = V_{Gd} = W_F + V_{G} &\quad \mbox{at} \quad y=y_1, \, x \in [x_2, x_3]\\
\nabla_{\nu}\phi = 0 & \quad \mbox{at the remaining part of the boundary}.
\end{array}
\end{eqnarray*}
Here $V_b$ is the bias voltage, $V_{S}$ is the source potential, $V_{D}$ is the source potential, $V_{Gu}$ is the upper gate potential, $V_{Gd}$ is the down gate potential. $V_{G}$ is the gate bias potential which is considered equal at both the gate contacts.  $\nabla_{\nu}$ denotes the normal derivative. 


 In graphene a sort of doping is induced by the electrostatic potential  \cite{Landauer} which leads to a shift of the Fermi energy. Assuming thermal equilibrium,  the initial carrier densities $n_0(x)$ and $ p_0(x)$ of electrons and holes respectively are related to the electric potential by
\begin{equation*}\label{EQ:pot_dens_rel}
\begin{alignedat}{3}
& n_0(x) && = \frac{2}{(2\pi)^2}\int \! f_{FD}(\mathbf{k}; e\phi(x,y_{gr})) \, d\mathbf{k}, && \quad x \in [x_1, x_4], \\
& p_0(x) && = \frac{2}{(2\pi)^2}\int \! f_{FD}(\mathbf{k}; -e\phi(x,y_{gr})) \, d\mathbf{k}, && \quad x \in [x_1, x_4],
\end{alignedat}
\end{equation*}
being $f_{FD}$ the Fermi-Dirac distribution
\begin{equation*}
f_{FD}(\mathbf{k};\varepsilon_F) = \left[1+\exp\left( \frac{\varepsilon(\mathbf{k})-\varepsilon_F}{k_B T} \right) \right]^{-1},
\end{equation*}
where $\varepsilon_F$ is the Fermi level (in pristine graphene $\varepsilon_F = 0$), $\varepsilon(\mathbf{k}) = \hbar v_F\vert \mathbf{k}\vert $ is the graphene dispersion relation (strictly valid around the Dirac points),
which is the same for electrons and holes (see \cite{BOOK:Jacoboni,Kittel,CaNe}), $\hbar$ is the reduced Planck constant and $v_F$ is the Fermi velocity. The crystal momentum of electrons and holes is assumed to vary over  $\mathbb{R}^2$.
The boundary conditions at the contacts are given by
\begin{eqnarray}
n(t, x_i) = n_0 (x_i), \quad p(t, x_i) = p_0 (x_i), \quad i = 1, 4. \label{Poisson_BC}
\end{eqnarray}
%

Altogether, in order to get the initial density profile as function of the electric potential, we must solve the following nonlinear Poisson equation 
\begin{equation}
\nabla\cdot(\epsilon\nabla\phi)=g(\phi), \label{IniPoiss1}
\end{equation}
augmented with the boundary conditions (\ref{Poisson_BC}), 
where
\begin{equation}
g(\phi(x,y)) =\left\lbrace
\begin{aligned}
&e \left[ n_0(\phi(x,y))-p_0(\phi(x,y))\right]/t_{gr} \qquad \mbox{if} \quad  (x, y) \in  [x_1, x_4]\times [y_2, y_3]\\
&0 \qquad\qquad\qquad\qquad\qquad\qquad\qquad\quad \mbox{otherwise}.
\end{aligned}
\right.
\end{equation}

\section{Mobility model}
\label{section:mobility}
We adopt the mobilities deduced in \cite{NaRo_DGFET} from a direct numerical simulation of the transport equations by using a DG scheme (the interested reader is referred to \cite{CoMajRo,ART:MajNaRo_CICP} for the details). The  behaviour is the same for holes on account of the symmetry between the hole and electron distributions. 


The low field mobility has the expression
\begin{eqnarray*}
\mu_0(n) &=& \tilde{\mu}_1 - \tilde{\mu}_0 \frac{\exp\left( -\frac{(\log(n/n_{ref})-m)^2}{2 \sigma^2} \right)}{\sqrt{2\pi} \sigma n/n_{ref}}
 \left( a\left(\frac{n}{n_{ref}}\right)^2+b\frac{n}{n_{ref}}+c\right), \label{lowmu}
\end{eqnarray*}
where the fitting parameters have been estimated by the least squares method as follows 
 $\tilde{\mu}_0$ = 0.2978 $\mu$m$^2$/V ps , $\tilde{\mu}_1 = $ 4.223  $\mu$m$^2$/V ps, $n_{ref}$ =  376.9 $\mu$m$^{-2}$, $m = - $0.2838, $\sigma$ = 2.216, $a$ =  4.820, $b$ = 68.34 and $c$ = 2.372.
 
%

The complete mobility model is given by (see also \cite{ART:NaRo_CAIM},\cite{MajMaRo})
\begin{equation}
\mu(E,n) = \frac{\mu_0(n)+\tilde{\mu}\left( \frac{E}{E_{ref}} \right)^{\beta_1}}{1+\left( \frac{E}{E_{ref}} \right)^{\beta_2}+\gamma\left( \frac{E}{E_{ref}} \right)^{\beta_3}}, \label{highmu}
\end{equation}
where $E_{ref}$, $\tilde{\mu}$, $\beta_1$, $\beta_2$, $\beta_3$ and $\gamma$ are fitting parameters. 
 
We have calculated the coefficients $E_{ref}$, $\beta_1$, $\beta_2$, $\beta_3$, $\gamma$ and $\tilde{\mu}$ by means of least square method for several value of the electron density $n_i$, obtaining the data reported in Tab.~\ref{TAB:High_field_mob_param}. The values of the density correspond to the Fermi energies 0.1, 0.2, 0.3, 0.4, 0.5 eV.
\begin{table}[htp]
\centering
\begin{tabular}{c|c|c|c|c|c|c}
$n_i$ & $E_{ref}$ & $\beta_1$ & $\beta_2$ & $\beta_3$ & $\gamma$ & $\tilde{\mu}$\\
\hline
4471.0 & 0.05265 & 1.034 & 2.135 & 1.059 & 14.53 & 12.78\\ 
15500.0 & 0.02126 & 0.4615 & 1.584 & 0.4276 & 21.77 & 33.3\\ 
33877.0 & 0.1096 & 1.344 & 2.52 & 1.457 & 4.595 & 6.579\\ 
59588.0 & 0.1776 & 2.304 & 3.099 & 1.335 & 1.395 & 1.251\\ 
92644.0  & 0.05047 & 1.988 & 2.661 & 1.109 & 0.8816 & 1.915
\end{tabular} 
\caption{High field mobility parameters.}
\label{TAB:High_field_mob_param}
\end{table}
In each interval $[n_i, n_{i+1}]$  a third degree polynomial interpolation has been adopted for the parameters $E_{ref}$, $\beta_1$, $\beta_2$, $\beta_3$, $\gamma$ and $\tilde{\mu}$ \cite{NaRo_DGFET}.

\section{Numerical simulations}
The GFET of Fig.~\ref{FIG:GMOSFET} has been intensively simulated. We have set 
the length 100 nm, the  width of the lower and upper oxide (SiO$_2$) 10 nm. The source and drain contacts are  long 50 nm. The lateral contacts are long 21 nm.
The two gate potentials are set equal in the simulated cases. We considered a mesh of 40  grid points along the $x$-direction and 23 grid points along the y-direction. In the graphene layer a single row of 40 nodes has been employed. 
In order to get the solution the following strategy has been adopted:
first the Poisson equation is solved by keeping the charge in the graphene layer equal to $n_0(x)$ and $p_0(x)$ for electron and holes, respectively; then the nonlinear Poisson problem (\ref{IniPoiss1}) is solved with an iterative scheme, by taking as initial guess the solution of  the previous step;
once the initial data for the electron and hole density have been determined, the full transient drift-diffusion-Poisson system is solved until the steady state is reached. One gets the stationary regime in about three picoseconds.
\begin{figure}[ht]
\centerline{\includegraphics[width=0.49\textwidth]{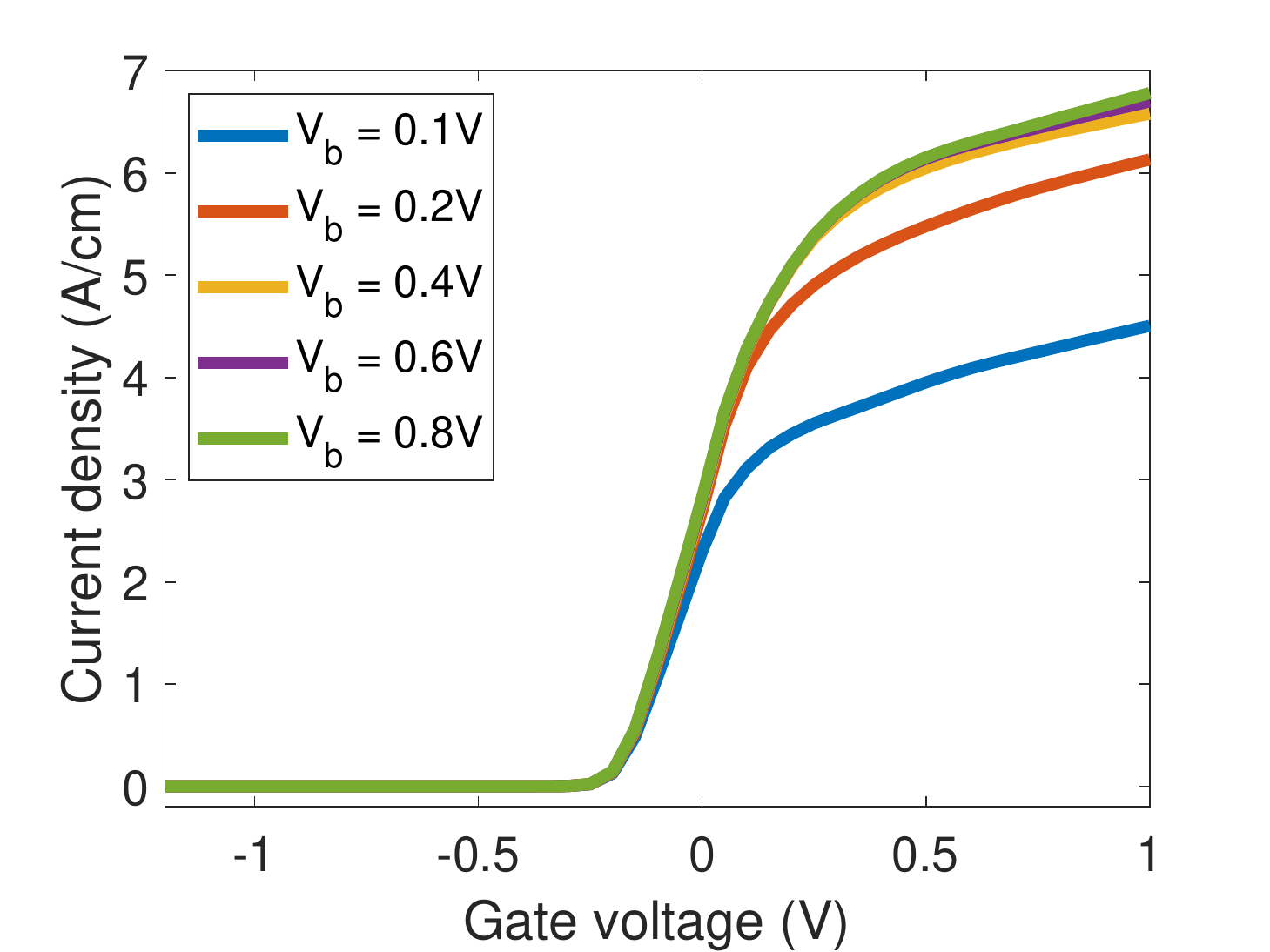}\includegraphics[width=0.49\textwidth]{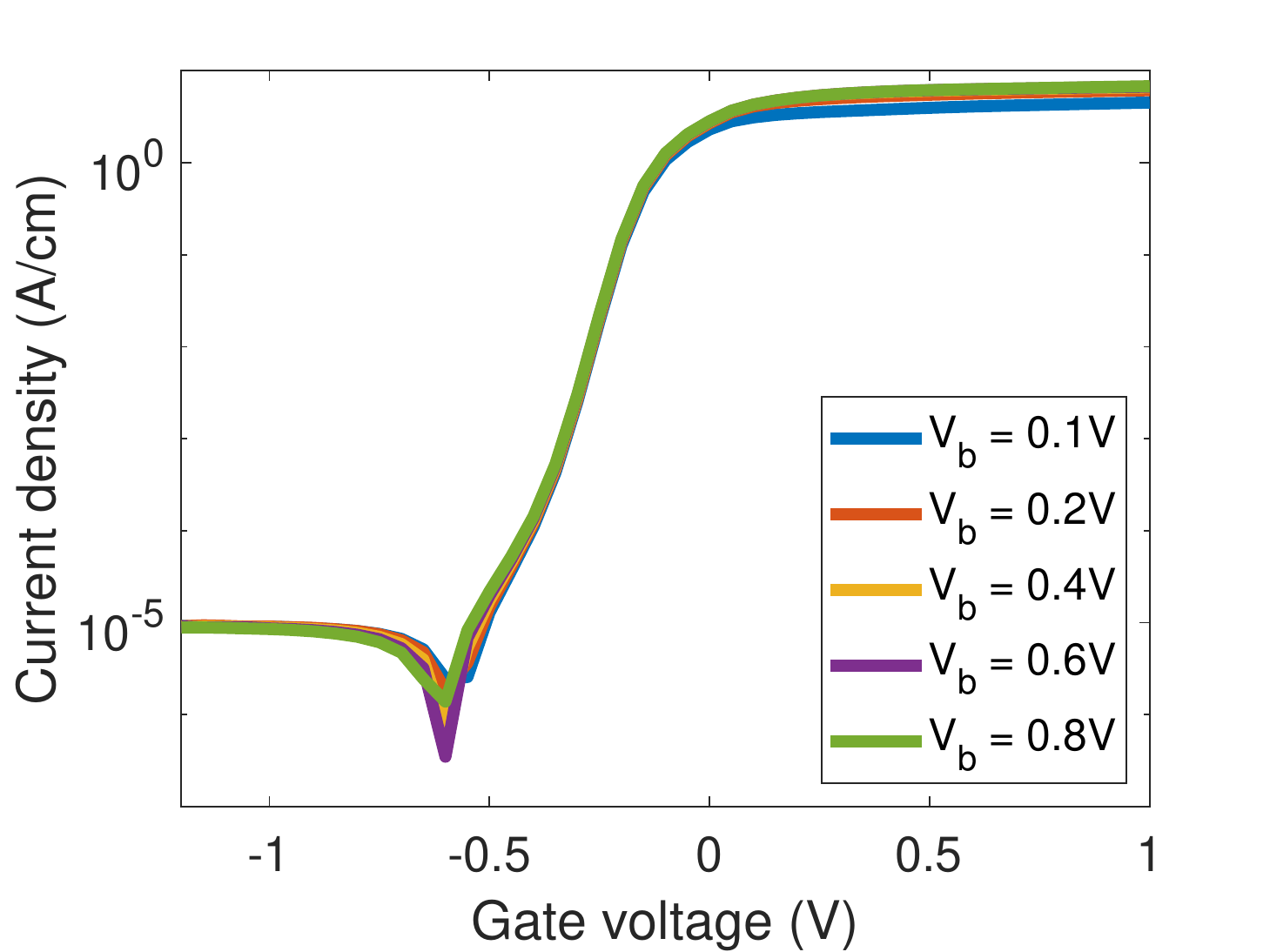}}
\caption{Total current versus gate voltage at fixed source-drain bias in a linear scale (left) and a semi-logaritmic scale (right).}\label{FIG:tot_curr_vs_VG}
\end{figure}
\begin{figure}[ht]
\centerline{\includegraphics[width=0.49\textwidth]{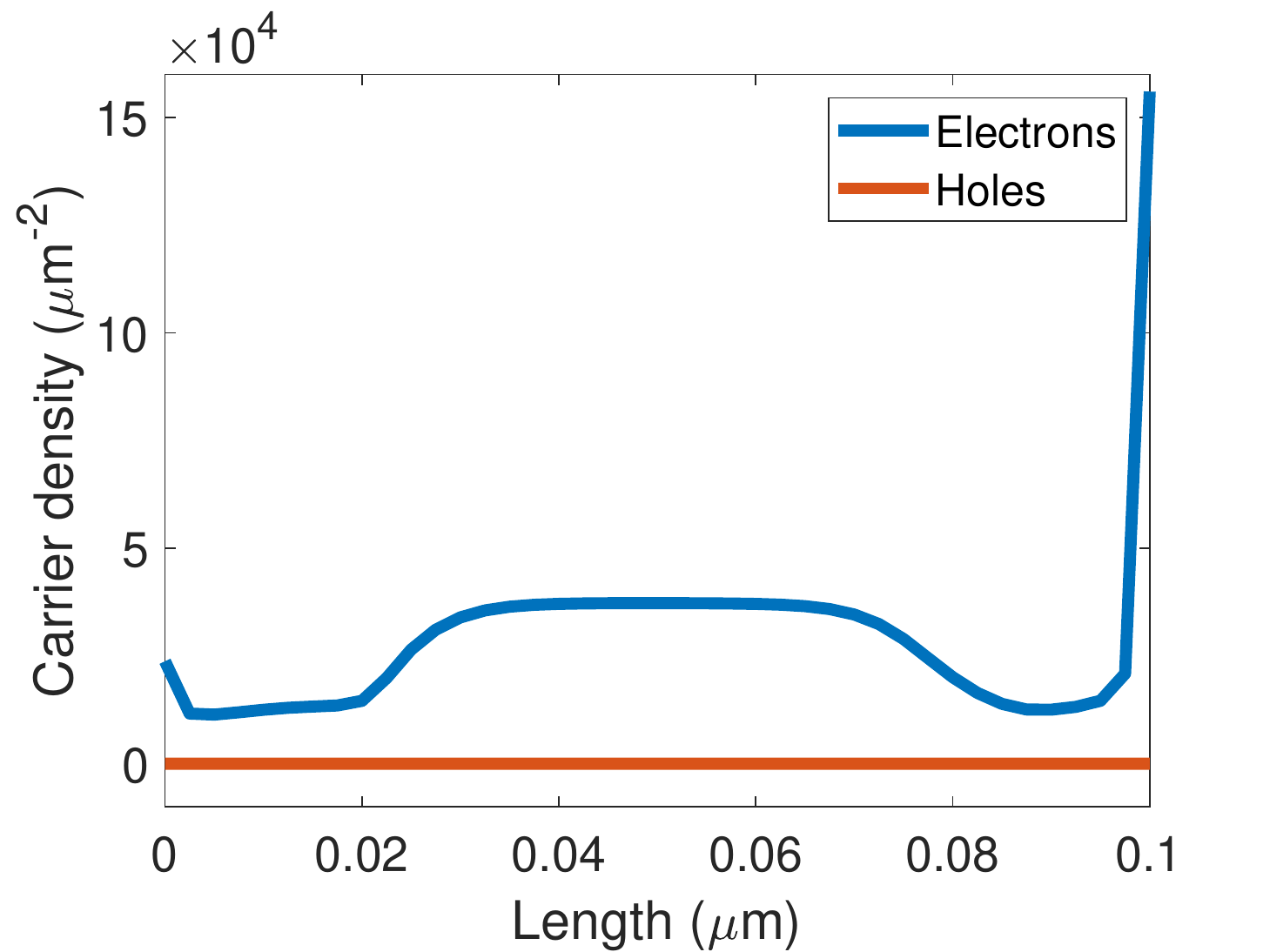}\includegraphics[width=0.49\textwidth]{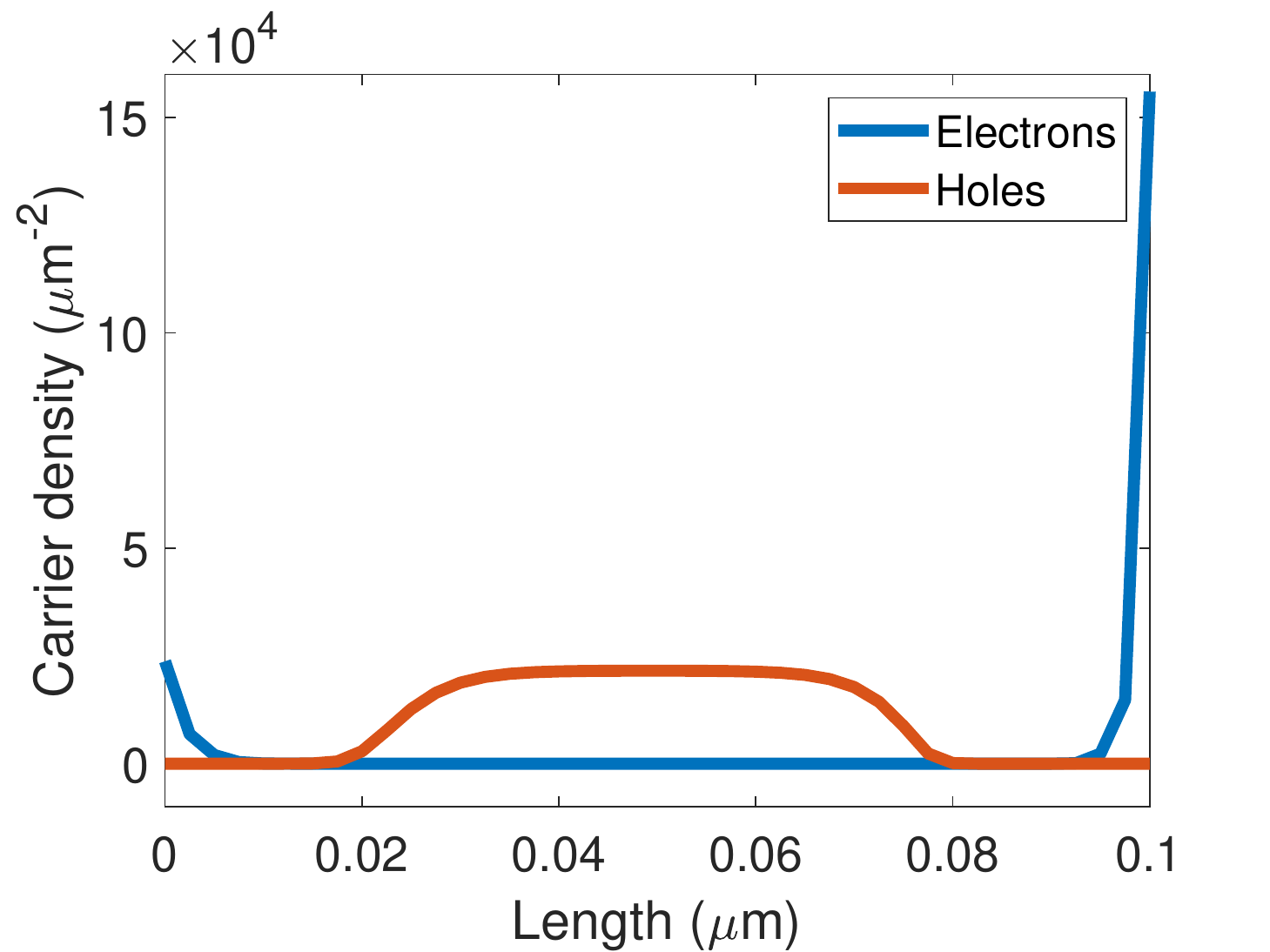}}
\caption{Carrier densities at $V_b$ = 0.4 V. Left: $V_G$ = 1 V (case on). Right:  $V_G =  -$1 V (case off). }\label{FIG:densities}
\end{figure}
\begin{figure}[ht!]
\centerline{\includegraphics[width=0.49\textwidth]{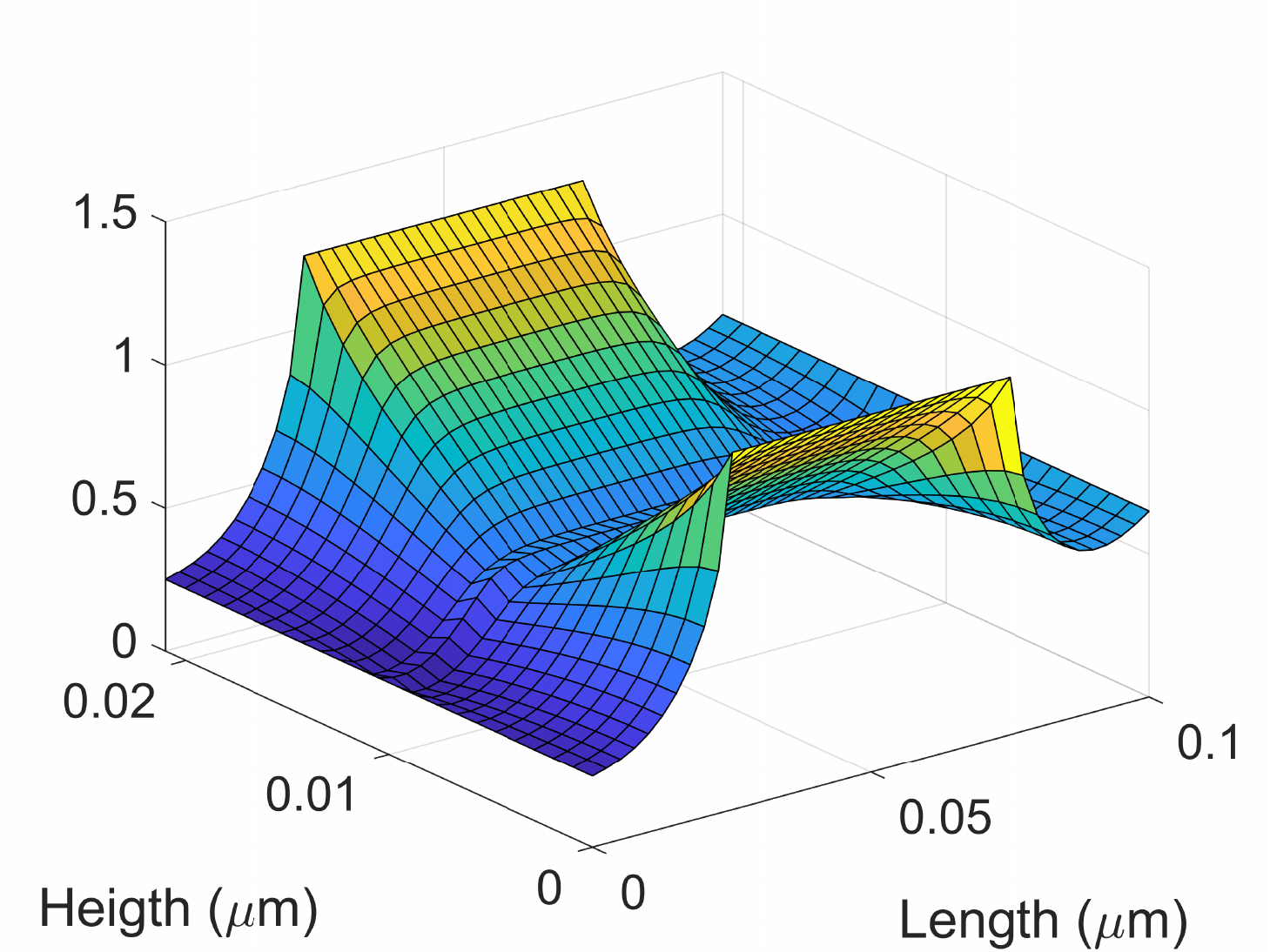}\includegraphics[width=0.49\textwidth]{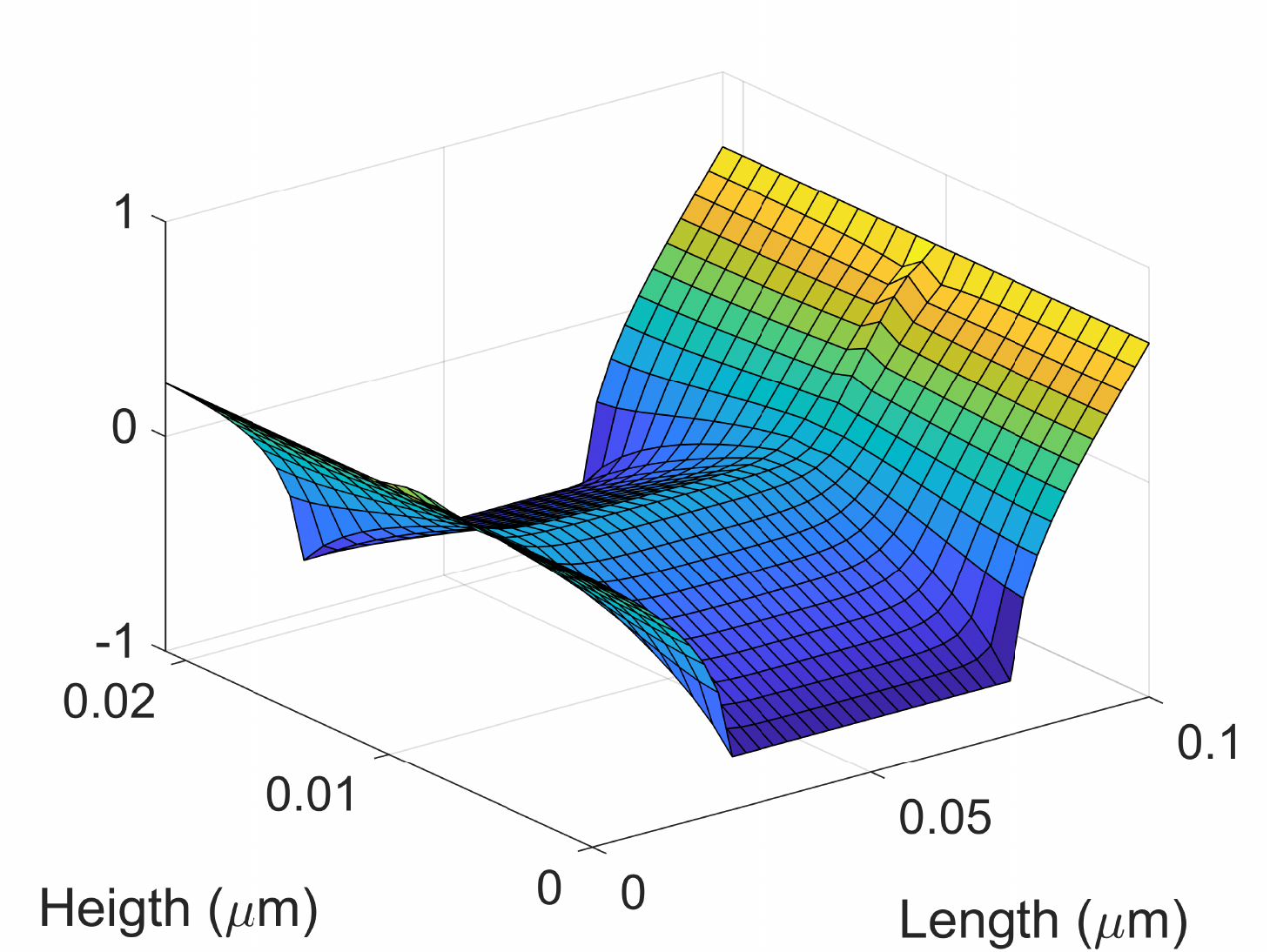}}
\caption{Electrostatic potential with $V_{b}$ = 0.4 V. On the left the case of $V_G = 1$ V. On the right  the case of $V_G = - 1$ V.}\label{FIG:potential}
\end{figure}
The characteristic curves are shown in  Fig.s~\ref{FIG:tot_curr_vs_VG}.
It is evident that an acceptable field effect transistor is obtained. By decreasing the gate voltage the current settles at a very low value. The plot in logarithmic scale shows a current-on/current-off ratio of five orders of magnitudo at least.  The minority charges are only slightly activated and the inversion gate voltage observed with other geometries \cite{Sch} is practically not present. From  Fig.s~\ref{FIG:densities} it is evident that the hole density remains negligible with respects to the majority one. The main reason for such an effect is the peculiar position and breadth of the source and drain contacts which create an electrostatic potential such that the electron total energy is always of constant sign.  
Devising a working structure has required a full 2D numerical solution of the Poisson equation. The common use of lumped 1D models for the electrostatic potential  has the drawback of hiding the effects related to the source and drain contact position and breadth. The simulations have been also performed adopting the  mobility model in \cite{Dor}. 

The results are qualitatively  in good agreement with those obtained by using the  mobility model of Sec. \ref{section:mobility}. Therefore we are rather confident that the efficient FET performance of the proposed device is  not an artifact related to the  
adopted specific mobility model but a general feature stemming from the chosen geometry. The proposed device overcomes the drawback related to the zero gap in pristine large area graphene and  represent a viable prototype for for the design of GFETs.
%
%
%
%
%
%
\section{Acknowledgements}
The authors  acknowledge the support from INdAM (GNFM). This work has been supported by the Universit\`{a} degli Studi di Catania, {\em Piano della Ricerca 2016/2018 Linea di intervento 2}.

\end{document}